\documentclass{article}

\PassOptionsToPackage{numbers, compress}{natbib}
\usepackage[preprint]{neurips_2026}

\usepackage[utf8]{inputenc} % allow utf-8 input
\usepackage[T1]{fontenc}    % use 8-bit T1 fonts
\usepackage{hyperref}       % hyperlinks
\usepackage{url}            % simple URL typesetting
\usepackage{booktabs}       % professional-quality tables
\usepackage{amsfonts}       % blackboard math symbols
\usepackage{nicefrac}       % compact symbols for 1/2, etc.
\usepackage{microtype}      % microtypography
\usepackage{xcolor}         % colors

\usepackage{graphicx}
\usepackage{multirow}
\usepackage{multicol}
\usepackage{amsmath}
\usepackage{verbatim} 
\usepackage{wrapfig}
\usepackage{booktabs}
\usepackage{tcolorbox}  
\usepackage{amsmath}
\usepackage{enumitem}
\usepackage{pifont}
\tcbuselibrary{listings,breakable}
\usepackage[ruled,vlined]{algorithm2e}

\title{HIDBench: Benchmarking Large Language Models for Host-Based Intrusion Detection}

\newcommand{\system}{\textsc{HIDBench}\ }
\newcommand{\systemc}{\textsc{HIDBench}}

\author{%
  Danyu Sun \\
University of California, Irvine\\
  \texttt{danyus2@uci.edu} \\
  \And
  Jinghuai Zhang \\
  University of California, Los Angeles \\
  \texttt{jinghuai1998@g.ucla.edu} \\
  \AND
  Yuan Tian \\
  University of California, Los Angeles \\
  \texttt{yuant@ucla.edu} \\
  \And
  Zhou Li \\
  University of California, Irvine\\
  \texttt{zhou.li@uci.edu } \\
}

\begin{document}

\maketitle

\begin{abstract}
 
Recent benchmark efforts have advanced the evaluation of large language models (LLMs) in cybersecurity, including tasks such as penetration testing and vulnerability identification. However, a critical cybersecurity task, namely intrusion detection from system logs, remains unexplored. In this work, we present a new benchmark to assess LLMs' capabilities in supporting host-based intrusion detection systems (HIDS). This task requires fine-grained reasoning over large-scale, noisy, and highly imbalanced system logs, where complex interactions between benign and malicious activities make reliable detection challenging. Our benchmark unifies three public system log datasets, DARPA-E3, DARPA-E5, and NodLink, and introduces a data construction pipeline that transforms raw host telemetry into LLM-compatible inputs, enabling systematic evaluation under realistic intrusion detection settings. Our evaluation of frontier LLMs reveals substantial performance gaps across datasets. While many models achieve high precision (often above 0.8) on simpler datasets, their performance degrades significantly as system logs become noisier and more complex, with MCC frequently dropping below 0.5 and false positive rates increasing sharply. We further analyze model behavior and identify distinct regimes, including conservative detectors with low false positive rates and over-sensitive models that generate excessive alerts. Overall, our results highlight that while LLMs show strong potential for HIDS, their effectiveness is highly sensitive to data complexity, and robust system design is essential for reliable deployment.

\end{abstract}

\section{Introduction}
\label{sec:intro+related}

Recent benchmark efforts have explored the evaluation of large language models (LLMs)  in cybersecurity. General benchmarks such as GLUE~\cite{glue} and MMLU~\cite{mmlu} assess broad language capabilities but do not capture domain-specific security challenges. More recent domain-specific benchmarks, including CyberSecEval~\cite{cyberseceval} and CyberBench~\cite{cyberbench}, focus on structured tasks such as vulnerability understanding and security-oriented NLP. Other efforts such as Cybench~\cite{cybench} and CyberGym~\cite{cybergym} evaluate LLM agents in CTF-style or vulnerability exploitation scenarios. 
In parallel,  specialized harnesses have been designed for cybersecurity tasks like penetration testing~\cite{xu2024autoattacker, deng2024pentestgpt}, vulnerability detection~\cite{meng2024large, liu2024exploring}, threat intelligence analysis~\cite{liu2023constructing, huang2024ctikg, sorokoletova2024towards}, and interpreting attack artifacts such as malicious scripts~\cite{deng2024raconteur}. 
These benchmarks/tasks focus on software implementations or offensive security. Yet, a benchmark is missing for a critical cyber-security task, namely intrusion detection on the log telemetry. As intrusion detection is performed more frequently than the other tasks by defenders and LLMs are demonstrating increasingly advanced reasoning capabilities, a natural question arises: can they support real-world cybersecurity operations that demand fine-grained analysis of massive, noisy telemetry streams? 

In this work, we systematically investigate their performance on Host-based intrusion detection System (HIDS)~\cite{liu2018host}, which presents a uniquely challenging evaluation setting. Unlike isolated text inputs, host telemetry consists of heterogeneous event streams, such as process executions, file accesses, network connections, and provenance dependencies, generated continuously at enterprise scale. Malicious evidence is rare and often fragmented across multiple events, requiring models to reason over weak and distributed signals under severe class imbalance and extended temporal contexts. Benchmark construction poses an additional challenge: widely used datasets such as DARPA-E3 ~\cite{DARPA3program} and DARPA-E5 ~\cite{darpa_e5} contain massive audit logs with realistic background noise, making naive LLM evaluation impractical. Direct prompting on raw logs is constrained by context limits and computational cost, while filtering to attack-only events removes the benign context necessary for realistic detection. 

Several works have proposed integrating LLMs into HIDS frameworks, including agent-based systems for explainable intrusion detection~\cite{idsagent}, knowledge-driven provenance-based detection systems~\cite{omnisecllm}, and hybrid pipelines~\cite{hybrids}. However, LLMs play a limited role in these works and their capabilities are not systematically assessed. Concurrently, Bilot et al. developed PIDSMaker~\cite{pidsmaker}, a benchmark containing log telemetry to evaluate the provenance-based intrusion detection systems (PIDS)~\cite{inam2023sok}. However, the customized design for PIDS does not match the strengths (e.g., built-in knowledge of cyber-attack patterns) and weaknesses (e.g., limited context window) of LLMs.

To address this gap, we introduce \systemc, a unified benchmark for evaluating LLMs in HIDS. \system integrates three public log datasets (DARPA-E3, DARPA-E5, and NodLink) through a coherent data pipeline that transforms raw host telemetry into LLM-compatible inputs. 
Rather than directly prompting on raw logs, we design an evaluation framework that organizes system events around attack-centric windows, preserving malicious activity alongside its surrounding benign context while keeping inputs within model context limits. We further construct structured representations of system behavior that enable models to reason over heterogeneous signals involving processes, files, and network interactions, directly targeting the core challenges of long-context reasoning, sparse attack signals, and class imbalance. HIDBench is designed as a model-agnostic harness: integrating a new LLM requires only specifying its API endpoint, with no changes to the data pipeline, prompts, and evaluation logic.

Our evaluation of 9 frontier LLMs reveals substantial and recurring performance gaps across datasets. While many models achieve high precision on simpler datasets, performance degrades sharply as logs become noisier and more complex, with Matthews Correlation Coefficient (MCC) frequently dropping below $0.5$ and false positive rates increasing substantially. Analysis of model behavior reveals distinct detection regimes, including conservative detectors that suppress alerts and over-sensitive models that generate excessive false positives. These findings underscore the importance of robust system design for reliable HIDS deployment and establish \system as a rigorous testbed for this underexplored but operationally critical task. Rather than introducing a new standalone intrusion detection algorithm, \system is designed as a unified benchmark and evaluation harness for systematically assessing LLM capabilities in host-based intrusion detection under realistic telemetry settings. Specifically, our main contributions are as follows:
\begin{itemize}
    \item \textbf{Benchmark.} We introduce \system, a unified and plug-and-play benchmark for evaluating LLMs in host-based intrusion detection. Our benchmark integrates three widely used datasets (DARPA-E3, DARPA-E5, and NodLink) into a standardized evaluation framework, enabling systematic and cross-dataset comparison.

    \item \textbf{System Design.} We design a principled pipeline that transforms large-scale, noisy, and highly imbalanced system telemetry into structured, LLM-compatible inputs. Our approach preserves both malicious activity and surrounding benign context, enabling realistic evaluation under LLM context constraints.

    \item \textbf{Comprehensive Evaluation and Insights.} We conduct an extensive evaluation of frontier LLMs and uncover consistent performance gaps across datasets. Our analysis reveals that high precision can be misleading under severe class imbalance, and identifies distinct model behaviors (e.g., conservative vs.\ over-sensitive detectors), highlighting key challenges for deploying LLM-powered HIDS.
\end{itemize}

\section{System Design}
\label{sec:system}

\begin{figure}[h]
\centering
\includegraphics[width=0.95\linewidth]{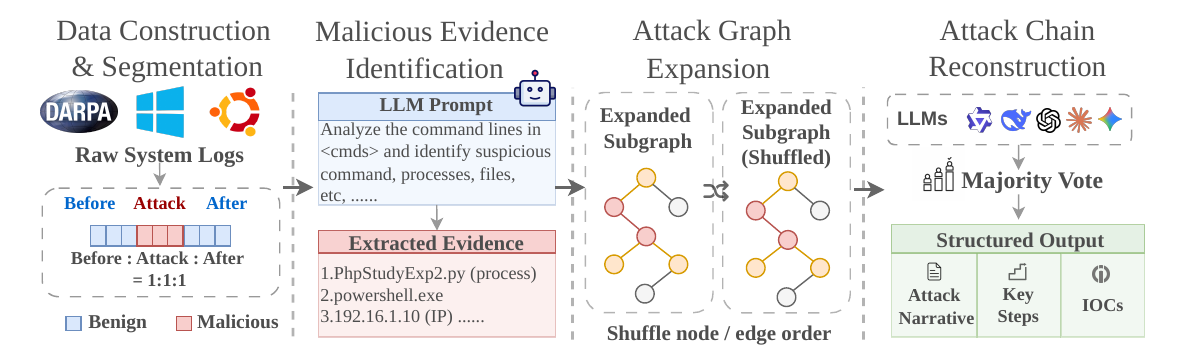}
\caption{Overview of the \system benchmark pipeline.}
\label{fig:pipeline}
\end{figure}

Figure~\ref{fig:pipeline} illustrates the workflow of our benchmark \systemc. It consists of four components: (i) \emph{Data Construction \& Segmentation}, which transforms high-volume, highly-imbalanced raw host logs into structured inputs suitable for evaluation by LLMs; (ii) \emph{Malicious Evidence Identification}, which extracts attack evidence using the LLM; (iii) \emph{Attack Graph Expansion}, which augments the identified evidence with its surrounding context; and (iv) \emph{Attack Chain Reconstruction}, which instructs the LLM to perform intrusion detection over the attack graph and produce structured outputs. Together, these components address the core challenges of scale, imbalance, and contextual complexity that make HIDS a particularly challenging setting for LLM.

\subsection{Data Construction \& Segmentation}
\label{subsec:datasplit}

%\textbf{Challenge.} 

%Existing HIDS datasets are separate and provide only raw host logs. 
Feeding the logs of an HIDS dataset directly into an LLM usually cannot generate meaningful results for two reasons. 
First, an HIDS dataset usually contains millions of events, far exceeding the input window of an LLM (e.g., 1M tokens). 
Second, though we can slice an HIDS dataset into smaller segments that fit in the LLM's input window, naive approaches like constant-interval sliding windows would severely degrade the input quality, because benign events often overwhelmingly dominate a window and cause class imbalance.

We carefully construct log segments by examining the ground-truth reports. To preserve execution semantics and system interactions, we retain identifiers, event attributes, and entity-level information (e.g., process paths, file paths, and network attributes) following prior provenance-based HIDS preprocessing practices~\cite{flash,orthrus}. We additionally remove redundant events to suppress noise. Detailed preprocessing descriptions are provided in Appendix~\ref{app:data_processing}. We first identify the attack interval that is documented by the report (e.g.,"/tmp/vUgefal" for DARPA E3 Cadets). Then, given a log stream $L$ with an attack interval $[t_s, t_e]$, we define the attack duration as $\Delta t = t_e - t_s$. We partition events into three temporally aligned segments: benign events before the attack, attack-period events, and benign events after the attack. Specifically, the benign context is selected using windows of the same duration as the attack interval, i.e., $[t_s-\Delta t, t_s)$ before the attack and $(t_e, t_e+\Delta t]$ after the attack. %\zl{do you select the benign events by time or number?}
The three segments are concatenated to form an attack-centric window that preserves both malicious activity and its surrounding benign context. Though the input window limits of the tested LLMs range from 131,072 (DeepSeek V3.2 and gpt-oss-120b) to 1,048,576 (Gemini-2.5-Flash) tokens, we found that none of the attack-centric windows exceed the limit.

\subsection{Malicious Evidence Identification (MEI)}
\label{subsec:mei}

%\textbf{Challenge.} 
An attack-centric window after the previous step
still contains a substantial amount of benign events that obscures malicious events. Directly prompting 
LLMs to detect the attack events over such inputs risks producing erroneous outputs due to the ``lost-in-the-middle'' effect~\cite{liu2024lost}.
To address this issue, we request the LLM to identify a 
compact set of high-confidence attack evidence $\mathcal{E}$ from the input events $E$. 
An HIDS dataset could contain various event types, but it is genuinely more difficult to classify certain types, e.g., network events with only IP addresses and flow types. Hence,  we focus on events with semantically rich signals such as command-line executions and process behaviors (e.g., execution of a suspicious process ``./gtcache'' via ``/bin/sh''
%\zl{an example of event}
). By anchoring LLM on the high-confidence attack evidence $\mathcal{E}$, we found that detection accuracy can be significantly improved. The prompt template for this step is described in Appendix~\ref{app:evidence_promt}.

\subsection{Attack Graph Expansion (AGE)}

%\textbf{Challenge.} 
High-confidence evidence identified by MEI captures salient malicious signals but lacks the surrounding context necessary for reasoning over multi-step attack behaviors. As an example, in the Browser Extension w/ Drakon Dropper attack from the DARPA-E3 THEIA dataset, the suspicious process ``./gtcache'' exhibited stealthy shell execution and repeated memory manipulation behaviors (e.g., EVENT\_MMAP and EVENT\_MPROTECT), but reconstructing the full attack chain required additional provenance context, including process lineage, file modifications, and external network communications. Without such context, benign intermediary processes and previously unseen network entities involved in the attack chain may be omitted from the investigation.

Therefore, we develop a lightweight context-building module to augment $\mathcal{E}$ from the previous step, inspired by prior provenance-based attack analysis works~\cite{kairos,flash,li2021survey}. Given the extracted evidence $\mathcal{E}$, we perform $k$-hop expansion to collect interacting entities and relations that capture contextual dependencies. The resulting subgraph is serialized into a structured representation for downstream reasoning. To reduce positional bias introduced by deterministic serialization, we randomly shuffle the order of nodes and edges before serialization while preserving graph structure and edge semantics. Appendix~\ref{app:shuffle_example} provides the pseudo-code of this procedure.

\subsection{Attack Chain Reconstruction (ACR)}
\label{subsec:acr}

Given the serialized subgraph produced by AGE, we prompt the LLM to generate two structured outputs: (i) an attack narrative describing the sequence of adversarial actions, and (ii) key indicators of compromise (IOCs) such as suspicious processes, files, and network connections. While IOCs can also be derived by traditional intrusion detection systems, attack narratives provide higher-level contextual explanations favored by human analysts. To ensure fair comparison, we use the same prompt template across all models without dataset-specific hints or handcrafted rules. Full prompt templates are provided in Appendix~\ref{app:attack_prompt}.

In this step, we also experiment with some post-processing techniques, including majority voting under test-time scaling~\cite{snell2024scaling} and iterative self-reflection~\cite{madaan2023self, shinn2024reflexion}. We found majority voting improves the detection results and include it as a default option in \systemc. Self-reflection produces mixed outcomes and we discuss this issue in Section~\ref{sec:abla}.
 
\section{Experiments and Results}
\label{sec:evalu}

\subsection{Experimental Setup}
\label{subsec:experisetup}

\textbf{Datasets.}
We evaluate HIDBench on three public host telemetry datasets: DARPA-E3~\cite{DARPA3program}, 
DARPA-E5~\cite{darpa_e5}, and NodLink-simulated data (NL-SD)~\cite{nodlink}, comprising 
nine sub-datasets in total. All datasets consist of host-level audit logs capturing 
endpoint system activities under realistic attack scenarios, with highly imbalanced 
distributions between malicious and benign events. Dataset statistics after ``Data Construction \& Segmentation'' described in Section~\ref{subsec:datasplit} are summarized 
in Table~\ref{tab:datasumm} in Appendix. The full dataset descriptions are provided in 
Appendix~\ref{app:data}.

\textbf{Ground-truth Labels.}
Labeling strategy significantly affects evaluation reliability, particularly given the severe class imbalance in host telemetry data. We adopt the labeling strategy that precisely models the attack activities.
For DARPA-E3/E5, we adopt the labeling strategy from Orthrus~\cite{orthrus-gt} and Velox~\cite{velox}, which strictly follows the official ground-truth documentation without additional expansion. Far fewer attack events are labeled compared to other more aggressive labeling strategies (e.g., multi-hop neighbors of annotated malicious nodes~\cite{flash,jia2024magic,wang2022threatrace}), which present a more challenging but realistic setting for intrusion detection.
For NL-SD, we use the ground-truth node labels directly provided by the dataset authors~\cite{nodlink-repo}.

\textbf{LLM Models.}
We evaluate a diverse set of frontier LLMs spanning multiple providers and development paradigms, including both proprietary and open-weight models.
Specifically, we include Claude-Opus-4.6~\cite{claude46}, Claude-Sonnet-4.5~\cite{claude45}, Claude-Sonnet-4~\cite{claude4}, GPT-5.2~\cite{gpt52}, GPT-4.1~\cite{gpt41} and Gemini-2.5-Flash~\cite{gemini25flash} as proprietary models and GPT-OSS-120B~\cite{gptoss120b}, DeepSeek-V3.2~\cite{deepseekv32} and Qwen3.6-Plus~\cite{qwen36plus} as open-weight models.
This selection covers a range of model families, scales, and training paradigms, enabling a comprehensive assessment of current LLM capabilities in HIDS.

\textbf{Evaluation Metrics.}
We evaluate model performance using precision, false positive rate (FPR), and Matthews Correlation Coefficient (MCC). Precision and FPR measure alert quality and false-alarm behavior, while MCC captures overall prediction quality under severe class imbalance. Detailed metric definitions are provided in Appendix~\ref{app:metrics}.

% We evaluate model performance primary using \emph{precision}, \emph{false positive rate (FPR) } and \emph{Matthews Correlation Coefficient (MCC)}, which together provide a comprehensive assessment under severe class imbalance. Precision measures the correctness of predicted alerts and is critical in intrusion detection, where false positives can overwhelm analysts and degrade system usability. FPR assesses the waste of human's time as every alert is expected to be examined by human analysts. Very low FPR is desired for an HIDS in the production environment. 
% MCC captures overall prediction quality by accounting for both positive and negative predictions, making it suitable for highly imbalanced settings. 
% %We additionally report the to characterize false-alarm behavior in practical deployment scenarios. 
% For FPR, we also characterize false-alarm behavior under different dataset and model combinations to gain a better understanding about when/why an error is caused.

%enables analysis of model behavior, allowing us to distinguish between conservative detectors with low false positives and over-sensitive models that generate excessive alerts.

\textbf{Data Contamination.}
A primary threat to validity is that the tested LLMs have seen the ground-truth labels of the public dataset and use them to answer prompts.
To mitigate such data contamination issue, our prompt templates do not expose dataset names, attack labels, or ground-truth annotations. 
Models only receive serialized provenance graphs and must reason over contextual system behaviors. 
We also manually inspected outputs and did not observe evidence of explicit ground-truth memorization. 
The substantial performance degradation on harder datasets such as DARPA-E5 further suggests that successful detection depends on contextual reasoning rather than memorized attack signatures.
%\zl{data contamination: need to specify that we verify the ground-truth is not memorized (or did you check every model?)}
%\zl{danyu/jinghuai, can you write something?}

% \begin{table}[t]
% \centering
% \scriptsize
% \setlength{\tabcolsep}{4pt}
% \caption{Dataset characteristics of our benchmark. Pos. Ratio denotes the percentage of malicious events/nodes.}
% \label{tab:dataset_characteristics}
% \small
% \begin{tabular}{llrrl}
% \toprule
% \textbf{Family} & \textbf{Host} & \textbf{\#Events} & \textbf{\#Mal.}  & \textbf{Mal:Benign} \\
% \midrule
% DARPA-E3 & CADETS & 39198  &  52 & 0.133\%\\
% DARPA-E3 & THEIA  &   121263 &  64  &  0.053\% \\
% DARPA-E3 & TRACE  &  106395 & 4  &  0.004\%\\
% DARPA-E5 & CADETS &   83323 &   18   & 0.022\%\\
% DARPA-E5 & THEIA  &  112741 &   56   & 0.050\% \\
% DARPA-E5 & TRACE &  61892 &  3   &  0.005\% \\
% NL-SD & HW17 & 481  &   16  & 0.034 \\
% NL-SD & HW20 & 1217  &  43 & 0.037 \\
% NL-SD & WIN10 &  2411 &  136  & 0.060\\
% \bottomrule
% \end{tabular}
% \end{table}

\subsection{Main Results}
We evaluate all models using precision, MCC, and FPR. Full per-model, per-dataset results are provided in Table~\ref{tab:main_results} in the Appendix. In the main text, we organize our findings under four research questions (\textbf{RQ1} to \textbf{RQ4}) highlighted below.

\subsection*{RQ1: How do dataset characteristics and complexity affect the performance of an LLM?}

\begin{figure}[h]
\centering
\includegraphics[width=0.89\linewidth]{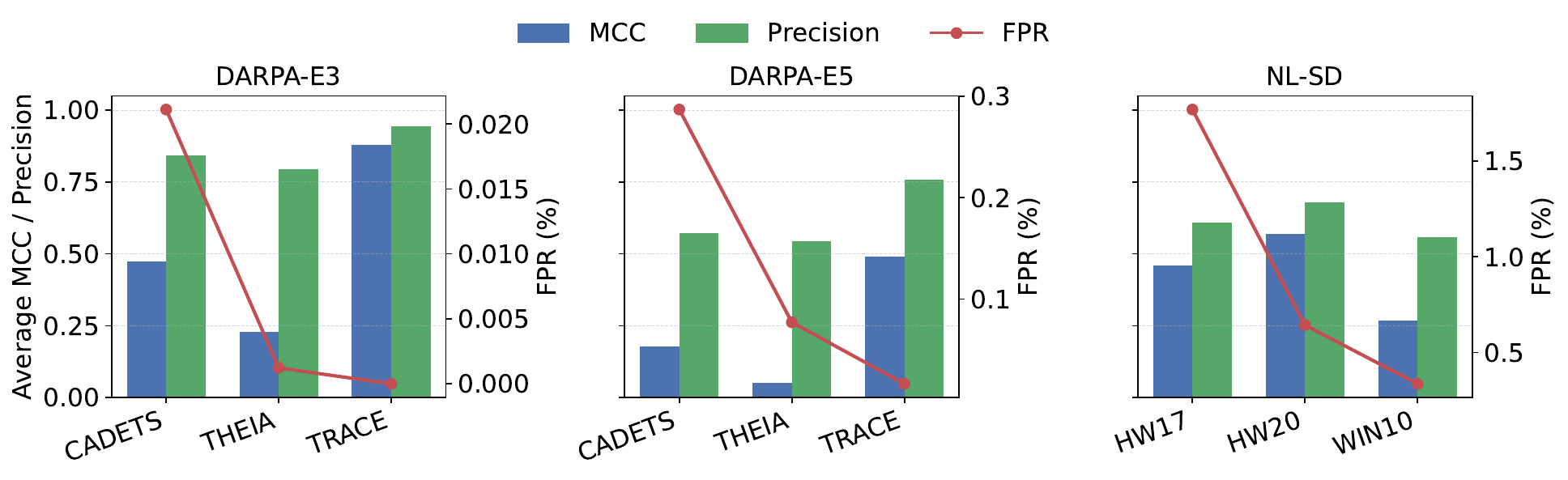}
\caption{Per-sub-dataset performance \textit{averaged} across all LLMs. Higher MCC and precision indicate better detection quality and alert accuracy, while lower FPR indicates fewer false alarms.
%\zl{which LLM are you using here?} \ds{this is all llm average resutl, should i extend them to all models?}
}
\label{fig:datadiff}
\end{figure}

Table~\ref{tab:datasumm} reveals substantial variation in both data volume and class imbalance across datasets.
The DARPA datasets are significantly larger, ranging from 39,198 to 121,263 events per sub-dataset, after data construction steps including attack-centric splitting and filtering based on unique event identifiers. These values reflect the size of the LLM-compatible inputs rather than the full raw logs. The datasets exhibit extremely low malicious-to-benign ratios, as low as 0.004\% in DARPA-E3 TRACE and 0.005\% in DARPA-E5 TRACE.
In contrast, NL-SD sub-datasets are considerably smaller (481--2,411 events) and exhibit higher malicious ratios, ranging from 3.4\% to 6.0\%.

%\zl{better to summarize the figures and numbers here}
%\zl{again, add some numbers, text is quite vague}

These structural differences translate directly into measurable performance variation, as shown in Figure~\ref{fig:datadiff} and Table~\ref{tab:main_results}. Across all evaluated LLMs, DARPA-E3 TRACE is consistently the easiest sub-dataset, achieving nearly perfect precision and MCC for most models (average precision: 1.000; average MCC: 0.879) with near-zero FPR. This suggests that attacks in TRACE expose highly distinguishable malicious behaviors that remain separable from benign background activity despite severe class imbalance. In contrast, DARPA-E3 THEIA exhibits substantially lower MCC despite relatively high precision. Averaged across models, THEIA achieves only 0.229 MCC, compared to 0.475 for CADETS and 0.879 for TRACE. For example, GPT-4.1 and GPT-OSS-120B both achieve perfect precision (1.000) on THEIA, yet their MCC scores remain low (0.250 and 0.216, respectively), indicating that identifying isolated suspicious events alone is insufficient for reconstructing the full attack chain. This behavior suggests that THEIA requires reasoning over long-range contextual dependencies between benign and malicious activities. CADETS presents an intermediate level of difficulty. Although its average MCC is substantially higher than THEIA, it consistently produces higher FPR than TRACE. Even small FPR values become operationally costly under the significantly larger event volume of provenance graphs.

Moving to DARPA-E5 and NL-SD, detection quality degrades substantially as dataset complexity increases. Compared to DARPA-E3, average precision on E5 drops from 0.839 to 0.237 on CADETS and from 0.781 to 0.071 on THEIA, while FPR increases significantly across models (e.g., GPT-4.1 rises from 0.02\% to 0.543\% FPR on CADETS). These results suggest that the noisier telemetry and more ambiguous attack behaviors in E5 substantially increase the difficulty of reliable attack reconstruction. NL-SD exhibits a different failure mode. While HW17 and HW20 maintain relatively strong average MCC values (0.458 and 0.570, respectively), WIN10 remains substantially harder despite its higher malicious ratio, achieving only 0.268 average MCC across models. This suggests that overlapping attack campaigns and interleaved attack chains introduce additional reasoning complexity beyond class imbalance alone. See Appendix~\ref{app:rq1_details} for additional dataset-specific analysis.

Taken together, these results reveal that detection difficulty is multidimensional and dataset-specific.
Class imbalance and background noise jointly explain the uniform performance degradation observed in DARPA-E5, while attack pattern visibility, rather than imbalance, is the primary driver of variation within DARPA-E3.
In NL-SD, neither imbalance nor noise is the bottleneck; instead, attack density and cross-platform behavioral ambiguity introduce qualitatively different challenges that current LLMs are ill-equipped to handle.
These findings suggest that no single dataset characteristic captures the full difficulty of HIDS evaluation, and that robust benchmarking must account for the interplay of multiple complexity factors simultaneously. 
\footnote{Recent non-LLM HIDS such as Orthrus~\cite{orthrus} and Velox ~\cite{velox}, report similar performance on DARPA-E3 and E5. For example, Orthrus and Velox achieve average precision of 0.66 and 0.85 on E3, and 0.43 and 0.12 on E5, respectively. However, the numbers do not necessarily reflect their real capability gaps due to different setups (e.g., the non-LLM HIDS have a training phase using the ``attack-free'' logs on the same machines, which are not used by \systemc). 
%However, these results are not directly comparable to ours due to differences in data construction, in particular our use of attack-centric splitting for LLM-based evaluation. While prior systems operate on full log streams with explicit graph structure, our benchmark constructs localized inputs that require LLMs to infer relationships from serialized representations.
}

%\zl{we could mention the results from traditional PIDS, e.g., Orthrus and Velox, compare with LLM}\textcolor{red}{Can we directly borrow results from our CCS paper?}\zl{maybe from their papers}

\subsection*{RQ2: How do frontier LLMs navigate the tradeoff between precision and false positive rate?}

\begin{figure}[h]
\centering
\includegraphics[width=0.89\linewidth]{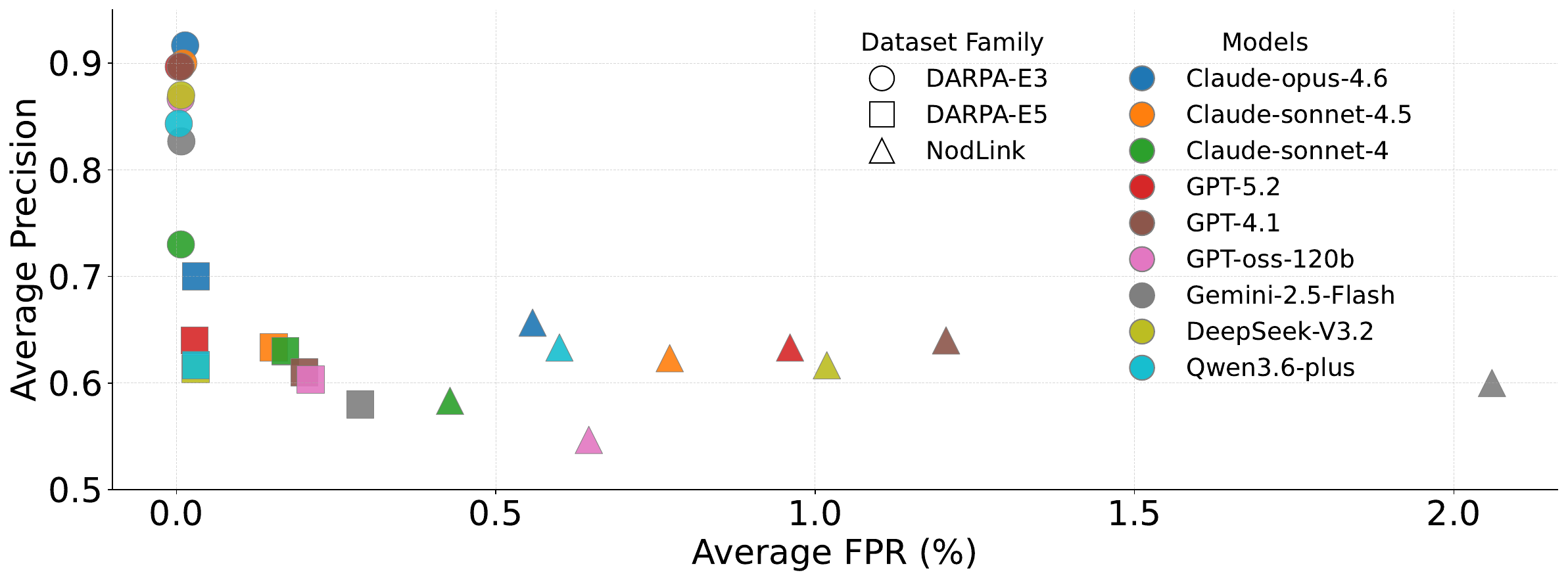}
\caption{Precision--FPR tradeoff across models and dataset families. Each point represents a model--dataset family pair, with marker shape indicating dataset family and color indicating model identity.}
\label{fig:pre_fpr}
\end{figure}

Figure~\ref{fig:pre_fpr} reveals a clear and consistent pattern: as dataset complexity increases, 
the precision--FPR tradeoff becomes increasingly difficult to maintain.

%\zl{need to add numbers in text. it's hard to follow your text}

On DARPA-E3, the simplest dataset family, most models cluster in the upper-left region of the plot, achieving both high precision and near-zero FPR. Averaged across models, DARPA-E3 achieves 0.893 precision with only 0.008\% average FPR, indicating that frontier LLMs can effectively suppress false alarms when attack behaviors remain distinguishable from benign activity. Claude-Opus-4.6 represents the strongest operating point in this regime, achieving approximately 0.92 average precision while maintaining only 0.01\% average FPR.

As dataset complexity increases, models spread more noticeably along the precision--FPR tradeoff curve. Compared to DARPA-E3, average precision on DARPA-E5 drops from 0.893 to 0.614, while average FPR increases from 0.008\% to 0.128\%. The tradeoff becomes most pronounced on NL-SD, where models separate into distinct behavioral regimes. Over-sensitive models aggressively flag suspicious behaviors at the cost of elevated FPR, while conservative models maintain lower FPR but suppress broader classes of attack indicators. For example, Gemini-2.5-Flash reaches approximately 2.1\% average FPR across NL-SD, substantially higher than other evaluated models. These results suggest that under noisier and more behaviorally heterogeneous environments, frontier LLMs increasingly diverge not only in overall capability, but also in their implicit detection strategies. Additional model-specific analysis of DARPA-E5 and NL-SD is provided in Appendix~\ref{app:rq2_details}.

%As dataset complexity increases in DARPA-E5, models spread noticeably along the precision--FPR tradeoff curve. Compared to E3, average precision drops from 0.893 to 0.614, while average FPR increases from 0.008\% to 0.128\%. Models that previously achieved similar operating behavior on E3 begin to diverge substantially under noisier telemetry. For example, GPT-4.1 reaches 0.543\% FPR on E5 CADETS despite only 0.02\% FPR on E3 CADETS, reflecting increased sensitivity to ambiguous benign behaviors.

%The tradeoff becomes most pronounced on NL-SD, where models separate into two distinct behavioral regimes. Over-sensitive models aggressively flag suspicious behaviors, achieving higher recall at the cost of elevated FPR, while conservative models maintain lower FPR but suppress broader classes of attack indicators. Gemini-2.5-Flash represents the most over-sensitive behavior, reaching approximately 2.1\% average FPR across NL-SD, substantially higher than all other models. In contrast, Claude-Opus-4.6 and Qwen3.6-plus maintain comparatively stable operating points, achieving around 0.64 average precision with average FPR below 1\%. These results suggest that as attack environments become noisier and more behaviorally heterogeneous, frontier LLMs increasingly differentiate not only in overall capability, but also in their implicit detection strategies.

%\zl{could talk about close vs open models in general, try to get some numbers}
We also observe a consistent gap between proprietary frontier models and open-weight/open-access models across datasets. Averaged over all sub-datasets, proprietary models achieve higher overall precision and substantially lower FPR. For example, on DARPA-E3, proprietary models achieve approximately 0.90 average precision compared to 0.82 for open models, while maintaining similarly negligible FPR. The gap becomes more pronounced on harder datasets such as DARPA-E5 and NL-SD, where open models exhibit substantially less stable operating behavior. For instance, GPT-OSS-120B and Gemini-2.5-Flash frequently drift toward high-FPR regimes under noisy telemetry, whereas Claude-Opus-4.6 and GPT-5.2 maintain comparatively stable precision--FPR tradeoffs. These results suggest that stronger frontier reasoning capabilities improve robustness under complex provenance analysis settings, particularly when attacks are heavily interleaved with benign behaviors.

Overall, these results demonstrate that the precision--FPR tradeoff is not a fixed property of an LLM, but strongly depends on dataset complexity and attack characteristics. While most frontier models achieve highly stable operating points on structured and lower-noise datasets such as DARPA-E3, their behaviors diverge substantially as background noise, attack density, and behavioral ambiguity increase. In particular, proprietary frontier models generally maintain more stable precision--FPR tradeoffs under complex settings, whereas open-weight/open-access models exhibit greater variance and are more likely to drift toward high-FPR regimes. These observations motivate a closer examination of the distinct behavioral patterns exhibited by different LLM families, which we explore in RQ3.

\subsection*{RQ3: Can frontier LLMs be categorized into distinct behavioral regimes, and what 
drives these differences?}

% To move beyond per-dataset comparisons, we characterize each model's false positive behavior 
% globally using two complementary metrics: its average FPR across all datasets $f_i^{\mathrm{avg}}$, 
% capturing overall false-alarm tendency, and its worst-case FPR $f_i^{\mathrm{max}}$, capturing 
% peak deployment risk. We define three behavioral regimes: \textbf{Conservative} models maintain 
% both low average and low worst-case FPR ($f_i^{\mathrm{avg}} < 0.25$, $f_i^{\mathrm{max}} < 1.0$); 
% \textbf{Over-sensitive} models exceed either threshold ($f_i^{\mathrm{avg}} \geq 0.50$ or 
% $f_i^{\mathrm{max}} \geq 2.0$); and \textbf{Balanced} models fall between these extremes.
% \zl{the motivation for this metric is still unclear. need better explanation}

To move beyond per-dataset comparisons, we characterize each model's false positive behavior globally across all datasets. Using only a single dataset-level FPR can be misleading because different provenance datasets exhibit substantially different event scales, attack densities, and background noise levels, leading to highly uneven FPR distributions and dataset-specific outliers. To better capture overall operational behavior, we therefore use two complementary metrics: the average FPR across all datasets $f_i^{\mathrm{avg}}$, which reflects a model's overall false-alarm tendency, and the worst-case FPR $f_i^{\mathrm{max}}$, which captures peak deployment risk under challenging conditions. Based on these metrics, we define three behavioral regimes. \textbf{Conservative} models maintain both low average and low worst-case FPR ($f_i^{\mathrm{avg}} < 0.25$, $f_i^{\mathrm{max}} < 1.0$). \textbf{Over-sensitive} models exceed either threshold ($f_i^{\mathrm{avg}} \geq 0.50$ or $f_i^{\mathrm{max}} \geq 2.0$), indicating unstable behavior under noisy conditions. \textbf{Balanced} models fall between these two extremes.

\begin{figure}[h]
\centering
\includegraphics[width=0.89\linewidth]{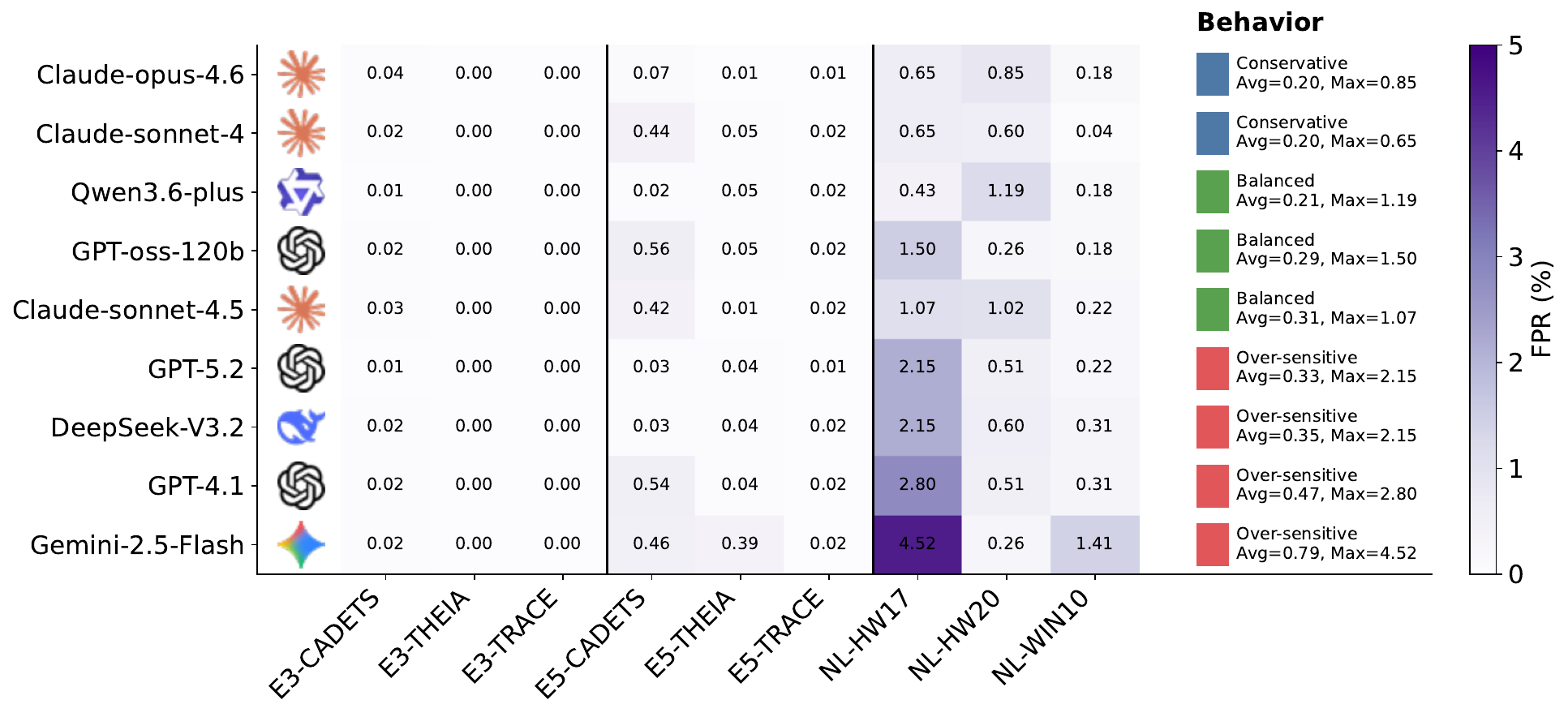}
\caption{Per-model FPR across all sub-datasets, with models sorted by average FPR. 
Color intensity reflects false-alarm severity, and behavior categories on the right 
are assigned based on both average and worst-case FPR, capturing overall tendency 
and peak deployment risk simultaneously.}
\label{fig:model_beha}
\end{figure}

Figure~\ref{fig:model_beha} reveals that this categorization is both meaningful and stable. Claude-Opus-4.6 and Claude-Sonnet-4 emerge as the most conservative models, maintaining consistently low FPR across the DARPA datasets while avoiding extreme false-positive spikes on NL-SD. At the other extreme, Gemini-2.5-Flash exhibits the most severe over-sensitivity, with FPR peaking at 4.52\% on NL-HW17 and remaining substantially elevated on WIN10 (1.41\%), despite near-zero FPR on DARPA-E3. GPT-4.1 and DeepSeek-V3.2 similarly fall into the over-sensitive regime, with GPT-4.1 reaching 0.54\% FPR on DARPA-E5 CADETS and both models exhibiting elevated false positives across multiple NL-SD settings. The balanced regime, occupied by Qwen3.6-Plus, GPT-OSS-120B, and Claude-Sonnet-4.5, represents models that avoid extreme false-alarm rates while maintaining moderate detection coverage.

A consistent pattern emerges across all regimes: FPR on DARPA datasets remains near zero for virtually all models, while NL-SD, particularly HW17, acts as the primary differentiator. This confirms that behavioral regime membership is not an intrinsic property of a model, but is revealed only under sufficiently challenging conditions. Models that appear behaviorally similar on simpler datasets diverge substantially under concurrent attacks and cross-platform behavioral ambiguity, underscoring the necessity of diverse and realistic benchmarks for meaningful model comparison.

%\zl{we need to talk about API cost, could be framed as overall costs, and fine-grained cost per dataset and IoC?}
%\textcolor{red}{Yes, I have asked Danyu to add it.}
%\zl{also time, reviewers seem care about it, see https://openreview.net/forum?id=oHEaIwPv9s}

\subsection*{RQ4: What is the cost and runtime overhead caused by LLMs?}

We further analyze the monetary cost of using LLMs for HIDS. Stronger frontier models such as Claude-Opus-4.6 and GPT-5.2 achieve better detection performance but incur substantially higher inference cost, evaluating a full dataset such as DARPA-E3 CADETS. In contrast, lightweight models such as Gemini-2.5-Flash and GPT-OSS-120B typically remain below \$0.1 per dataset, though often with reduced robustness on more challenging datasets. Runtime overhead follows a similar trend, with larger reasoning-oriented models requiring substantially longer inference time; however, even the slowest models complete inference within approximately one minute per sample. Detailed per-model cost and runtime statistics are provided in Appendix~\ref{app:cost}.

\section{Advanced Capabilities on HIDS: Reasoning, Scaling, and Reflection}
\label{sec:abla}

The main results reveal substantial performance gaps across datasets, raising a natural 
follow-up question: can these gaps be narrowed through more sophisticated inference 
strategies, without modifying model parameters? We investigate three complementary 
directions: explicit reasoning, test-time scaling via majority voting, and iterative 
self-reflection, probing the extent to which inference-time strategies can compensate 
for the inherent difficulty of HIDS.

%We conduct ablation studies to understand the impact of key components in our pipeline, focusing on reasoning strategies, test-time scaling, and iterative refinement. We organize the analysis around the following research questions:

% \paragraph{\textcolor{red}{RQ0: Graph-structured input vs. plain-text input for LLMs.}}

\textbf{How much does LLM reasoning rely on inferred versus explicit evidence?} To understand whether LLMs rely on overtly suspicious signals or contextual reasoning, we analyze reasoning traces produced by DeepSeek on DARPA-E3 CADETS. We first use GPT-5.4 for initial IOC categorization and then manually verify all classifications. We divide IOCs into two categories: \textit{explicit} IOCs, corresponding to directly observable malicious artifacts, and \textit{inferred} IOCs, which require contextual reasoning over co-occurring events and provenance relationships. Across all analyzed attack cases, inferred IOCs consistently outnumber explicit ones, suggesting that the model actively constructs contextual evidence chains beyond directly observable malicious indicators. Additional quantitative statistics and case studies are provided in Appendix~\ref{app:reasoning}.

% \begin{wraptable}{r}{0.45\linewidth}
% \centering
% \scriptsize
% \setlength{\tabcolsep}{3pt}
% \caption{\# of explicit and inferred IOCs.}
% \label{tab:reasoning_counts}
% \begin{tabular}{lccc}
% \toprule
% Attack & Qual. & \#Exp. & \#Inf. \\
% \midrule
% DARPA-E3 CADETS-A1 & HIGH & 4  & 6 \\
% DARPA-E3 CADETS-A2 & MEDIUM & 7  & 14 \\
% DARPA-E3 CADETS-A3 & MEDIUM & 7  & 15 \\
% DARPA-E3 CADETS-A4 & MEDIUM & 11 & 6 \\
% % \midrule
% % Avg(H) &  & 4.0  & 6.0 \\
% % Avg(M) &  & 8.3  & 11.7 \\
% \bottomrule
% \end{tabular}
% \end{wraptable}

As shown in Table~\ref{tab:reasoning_counts}, inferred IOCs consistently outnumber explicit ones 
across all four attack cases, with the gap most pronounced in medium-quality cases. This 
suggests that LLMs do not simply pattern-match on obvious indicators, but actively construct 
evidence chains that extend beyond directly observable signals. Representative examples in 
Table~\ref{tab:reasoning_examples} illustrate this: while explicit IOCs capture artifacts such as 
suspicious executables and external IPs, inferred IOCs surface subtler relationships such as 
system services and configuration files implicated through multi-step behavioral context. Further case analysis is provided in Appendix~\ref{app:reasoning}.

%This reliance on inferred evidence has a dual implication. On one hand, it enables detection of stealthy attacks whose signals are too weak or fragmented to be identified in isolation. On the other hand, it introduces a risk of overgeneralization, where benign processes that co-occur with malicious activity are incorrectly implicated. The balance between these two effects determines whether reasoning ultimately helps or hurts detection reliability.

\begin{table}[h]
\centering
\caption{Representative explicit and inferred IOCs of DARPA-E3.}
\label{tab:reasoning_examples}
\small
\begin{tabular}{lcll}
\toprule
\textbf{Case} & \textbf{Quality} & \textbf{Explicit IOCs} & \textbf{Inferred IOCs} \\
\midrule
CADETS-A1 & High &
\texttt{vUgefal, 61.167.39.128, /tmp} &
\texttt{/etc, /var, /dev} \\

CADETS-A2 & Medium &
\texttt{nginx, /etc/passwd, /tmp/grain} &
\texttt{76.56.184.25, rc.conf} \\

CADETS-A3 & Medium &
\texttt{minions, XIM} &
\texttt{/tmp/minions, sshd} \\

CADETS-A4 & Medium &
\texttt{pEja72mA, /tmp/*.so} &
\texttt{128.55.12.122, sshd} \\
\bottomrule
\end{tabular}
\end{table}

\begin{figure}[h]
\centering
\includegraphics[width=1.0\linewidth]{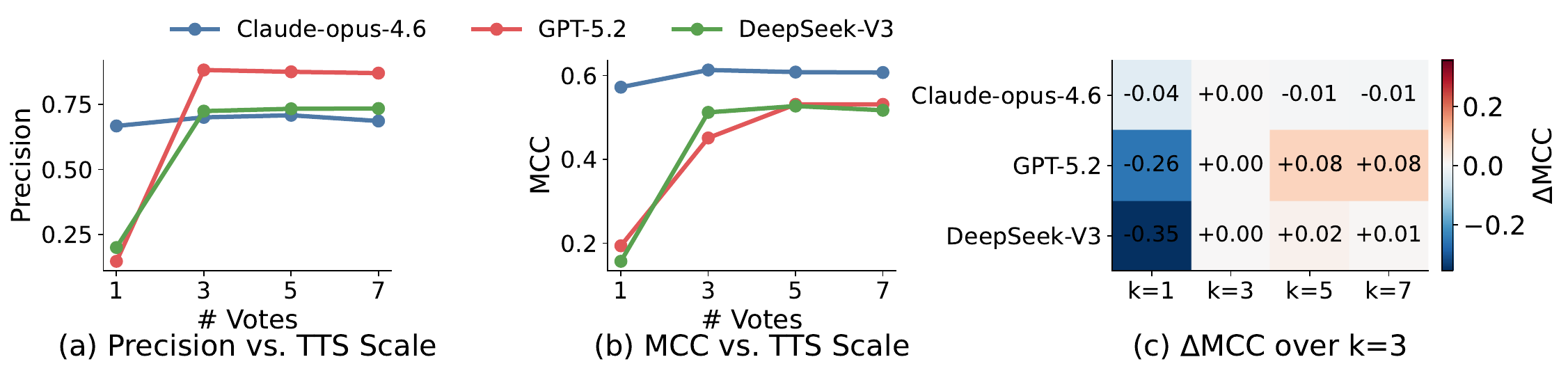}
\caption{
% Model behavior categorized by false positive rate across all datasets. Each cell reports the FPR of a model on a specific dataset. Models are sorted by average FPR, and the rightmost labels indicate behavior categories based on both average and worst-case FPR.
{Effect of test-time scaling on detection performance. (a) Precision and (b) MCC versus the number of voting samples for three representative LLMs. (c) $\Delta$MCC relative to the $k=3$ baseline, showing that gains largely saturate beyond $k=3$. Stronger models (Claude-Opus-4.6) remain stable even at $k=1$, whereas weaker models benefit more from aggregation.}
}
\label{fig:tts}
\end{figure}

\textbf{Does test-time scaling improve detection performance?} We evaluate majority voting under test-time scaling with sample sizes $k \in \{1,3,5,7\}$. As shown in Figure~\ref{fig:tts}, increasing $k$ from 1 to 3 yields the largest performance improvement across models, while larger values provide diminishing returns. Stronger models such as Claude-Opus-4.6 remain relatively stable even at $k=1$, whereas weaker or more variable models benefit substantially from aggregation. See Appendix~\ref{app:tts} for details.

%\zl{need to cite some works to justify why this is a good idea for test-time scaling, for our problem, and whether it's mainstream approach} \ds{add it the system design?}

% Recent studies show that test-time scaling can improve LLM reasoning performance without modifying model parameters~\cite{snell2024scaling,wu2024inference}. Approaches such as self-consistency and majority voting improve reasoning robustness by aggregating multiple independent generations~\cite{wang2022selfconsistency}. This strategy is particularly relevant for intrusion detection, where noisy and heterogeneous security logs can lead to unstable reasoning behavior.

% We evaluate majority voting with sample sizes $k \in \{1, 3, 5, 7\}$, where $k=3$ serves 
% as the default configuration in our benchmark. As shown in Figure~\ref{fig:tts}, the 
% transition from $k=1$ to $k=3$ yields the most substantial gains: both GPT-5.2 and 
% DeepSeek-V3 recover from near-random performance at $k=1$ to competitive levels at $k=3$, 
% confirming that a small ensemble already provides meaningful stabilization over single-sample 
% inference. Claude-Opus-4.6, by contrast, exhibits stable performance even at $k=1$, 
% suggesting that stronger base models are less reliant on voting to produce reliable predictions. Beyond $k=3$, scaling provides negligible additional benefit. As shown in panel (c), 
% $\Delta$MCC remains close to zero at $k=5$ and $k=7$ for all models, indicating that the 
% default setting of $k=3$ captures most of the available gain while avoiding unnecessary 
% computational overhead.

% \begin{table*}[t]
% \centering
% \small
% \begin{tabular}{lcccccccc}
% \toprule
% & \multicolumn{2}{c}{\# Vote = 1} 
% & \multicolumn{2}{c}{\# Vote = 3}
% & \multicolumn{2}{c}{\# Vote = 5}
% & \multicolumn{2}{c}{\# Vote = 7} \\
% \cmidrule(lr){2-3}
% \cmidrule(lr){4-5}
% \cmidrule(lr){6-7}
% \cmidrule(lr){8-9}
% Model & Pre & MCC & Pre & MCC & Pre & MCC & Pre & MCC \\
% \midrule
% Claude-opus-4.6  & 0.667 & 0.572  & 0.700  &  0.613 & 0.708  & 0.608 &  0.686 &  0.607 \\
% % Claude-sonnet-4.5  &  &   &   &   &   &  &   &   \\
% % Claude-sonnet-4  &  &   &   &   &   &  &   &   \\
% GPT-5.2    & 0.147 &  0.194 & 0.882  & 0.451  & 0.875  & 0.531 &   &   \\
% % GPT-4.1    &  &   &   &   &   &  &   &   \\
% % GPT-OSS-120B    &  &   &   &   &   &  &   &   \\
% % Gemini-2.5-Flash &   &    &   &    &    &    &   &    &    &    &    &   \\
% DeepSeek-V3 & 0.200 &  0.157 & 0.724 & 0.512 &   0.733 &  0.527 &   &   \\
% % Qwen3-235B-A22B &   &    &   &    &    &    &   &    &    &    &    &   \\
% \bottomrule
% \end{tabular}
% \caption{Ablation study of majority voting for test-time scaling on E3-CADETS. 
% We vary the number of votes (k = 1, 3, 5, 7) and report precision (Pre) and MCC.}
% \label{tab:majority_vote_ablation}
% \end{table*}

\textbf{Does iterative self-reflection improve detection performance?} Recent studies suggest that self-reflection can improve reasoning consistency and error correction in LLMs~\cite{kamoi2024can,pan2023automatically}. We evaluate two self-reflection strategies combined with majority voting: \textit{Ref-then-Agg}, where each sample is refined before aggregation, and \textit{Agg-then-Ref}, where self-reflection is applied once after aggregation.

%Overall, neither strategy consistently outperforms the baseline across all models, suggesting that the effectiveness of self-reflection is highly model-dependent. While some models benefit from reflection after aggregation, others exhibit reduced MCC despite improved precision, indicating a tradeoff between conservative prediction behavior and attack coverage. Detailed ablation results and model-specific analysis are provided in Appendix~\ref{app:self_reflection}.

Overall, neither strategy consistently outperforms the baseline across all models, suggesting that the effectiveness of self-reflection is highly model-dependent. For example, GPT-5.2 benefits substantially from Ref-then-Agg, improving MCC from 0.510 to 0.675, while DeepSeek achieves its best performance under Agg-then-Ref, improving MCC from 0.428 to 0.553. In contrast, Claude exhibits reduced MCC under both reflection strategies despite slight precision improvements, indicating a tradeoff between conservative prediction behavior and attack coverage. Detailed ablation results and additional analysis are provided in Appendix~\ref{app:self_reflection}.

\section{Conclusion}
\label{sec:conclu}

%\zl{need a paragraph about limitations and future work}

We present \systemc, the first benchmark for evaluating LLMs in host-based intrusion detection, unifying three public provenance graph datasets under a common evaluation framework. Our evaluation shows that LLM performance is highly sensitive to dataset complexity. Detection quality degrades substantially as background noise, class imbalance, and attack density increase, exposing distinct failure modes that simpler benchmarks cannot capture. Models further diverge into conservative and over-sensitive behavioral regimes under more challenging settings. Inference-time strategies such as test-time scaling and self-reflection provide targeted improvements, but no universal solution across models and datasets. These findings suggest that current LLMs are better suited as a triage and investigation layer than as primary detectors. Reliable deployment will require advances in both model robustness and hybrid system design beyond current inference-time techniques.
% These findings suggest that while LLMs show strong potential for HIDS, reliable deployment requires advances in both model robustness and system design beyond current inference-time techniques.

\textbf{Limitations and Future Work.} 
\system does not explore the entire design space for the LLM harness, such as a retrieval-augmented generation (RAG) module, which requires dataset-specific tuning. The generation of attack-centric windows relies on pre-defined heuristics and the detection results may vary with different parameter choices. \system is not a real-time detector as it waits for the entire subset processed before running the detection logic.
%and focuses only on inference-time evaluation without model adaptation. 
Future work includes extending the benchmark to streaming settings, incorporating more real-world telemetry datasets, and exploring more design choices for the harness.

\newpage

% \bibliographystyle{plainnat}
% \bibliography{main}

\bibliographystyle{IEEEtran}
\bibliography{main}

% \newpage
% \input{checklist.tex}

% \newpage
\appendix

\section{Impact Statement}
\label{app:impact}
This paper presents \system, a benchmark for evaluating large language models in 
host-based intrusion detection. Our work is intended to advance the scientific understanding 
of LLM capabilities in cybersecurity, with the goal of supporting the development of more 
reliable and robust intrusion detection systems.

We acknowledge that benchmark datasets and evaluation pipelines of this kind could 
potentially be misused to identify weaknesses in detection systems or to craft evasion 
strategies. However, we believe that the benefits of open evaluation and transparent 
model comparison substantially outweigh these risks. All datasets used in this work are 
publicly available and have been widely used in prior academic research. Our benchmark 
does not introduce new attack capabilities or provide operational attack tooling.

We hope that \system will serve as a foundation for future research into more robust 
LLM-based security systems, and encourage the community to use these resources 
responsibly in pursuit of stronger defensive capabilities.

\section{System Design Details}

\subsection{Data Pre-processing Details}
\label{app:data_processing}

To guarantee input quality, we retain fields that capture execution semantics and system interactions, following standard normalization practices in prior HIDS work~\cite{flash,orthrus}. Specifically, we preserve three categories of information:

\begin{itemize}[leftmargin=*]
    \item \textbf{Identifiers:} subject and object identifiers used to preserve provenance relationships.
    
    \item \textbf{Event attributes:} event type, command line, timestamp, and related execution metadata.
    
    \item \textbf{Entity information:} process paths, file paths, and network attributes such as IP addresses and ports when available.
\end{itemize}

When entity-level information is unavailable, we rely on identifiers and event attributes only. We additionally remove redundant or repeated events to reduce noise and avoid over-representation of frequent benign activities in serialized provenance graphs.

\subsection{Graph Serialization Shuffle Example}
\label{app:shuffle_example}
To reduce positional bias introduced by deterministic graph serialization, we randomly permute the order of nodes and edges before converting the provenance graph into textual input for the LLM, as summarized in Algorithm~\ref{alg:graph_shuffle}. The underlying graph structure and edge semantics remain unchanged; only the serialization order is modified.

\begin{algorithm}[h]
\caption{Graph Serialization Shuffle}
\label{alg:graph_shuffle}
\KwIn{Provenance graph $G=(V,E)$}
Randomly permute node order in $V$\;
Randomly permute edge order in $E$\;
Serialize shuffled $(V,E)$ into textual graph format\;
\Return serialized graph representation\;
\end{algorithm}

\section{ Additional Experiment Details}
\label{sec:app}

%\zl{could add more prompt and response examples here}

\subsection{Dataset Description}
\label{app:data}
\textbf{DARPA-E3/E5.}
The DARPA-E3 and DARPA-E5 datasets are collected under the DARPA Transparent Computing 
(TC) program, and are widely used in host-based intrusion detection research~\cite{inam2022sok}.
Red-team attackers simulate real-world attack scenarios within enterprise environments, 
while host-level audit logs are collected from endpoint systems for forensic analysis.
The logs contain both malicious activities and substantial benign background behavior, 
resulting in realistic and highly imbalanced data distributions.
Both datasets are released into the public domain by DARPA to facilitate research, 
with no restrictions on use~\cite{DARPA3program}.
We evaluate on three sub-datasets per version: CADETS, THEIA, and TRACE.

\textbf{NodLink (NL-SD).}
The NL-SD dataset~\cite{nodlink} is generated from a simulated enterprise environment 
consisting of multiple hosts with heterogeneous operating systems. It includes realistic 
attack scenarios such as Apache Struts2-046 along with normal system operations, and 
covers both Linux and Windows environments across three sub-datasets: HW17, HW20, and WIN10.
The dataset is publicly released under a Research-Only MIT License, permitting academic 
and educational use~\cite{nodlink-repo}.

\begin{table}[h]
\centering
\caption{Dataset Summary.}
\small
\label{tab:datasumm}
\begin{tabular}{lrrl}
\toprule
\textbf{Dataset} & \textbf{\#Event} & \textbf{\#Malicious} &\textbf{Malicious:Benign} \\
\midrule
DARPA-E3 CADETS & 39,198 & 52 & 1:753\\
DARPA-E3 THEIA  & 121,263 &64 & 1:1894\\
DARPA-E3 TRACE  & 106,395 & 4 & 1:26598\\
DARPA-E5 CADETS & 83,323 & 18 & 1:4628\\
DARPA-E5 THEIA  & 112,741 & 56 & 1:2012\\
DARPA-E5 TRACE  & 61,892 & 3 & 1:20630\\
NL-SD HW17   & 481 & 16 & 1:29\\
NL-SD HW20   & 1,217 & 43 & 1:27\\
NL-SD WIN10  & 2,411 & 136 & 1:17\\
\bottomrule
\end{tabular}
\end{table}

\subsection{Running environment}
\label{app:comresource}
All running and evaluation were conducted on a server running Ubuntu 22.04, equipped with a 3.8GHz 16-core Intel(R) Xeon(R) Gold 5222 CPU, 768 GB of memory, and an NVIDIA A6000 GPU with 48GB of memory.

\subsection{Evaluation Metrics}
\label{app:metrics}

\textbf{Precision} measures the correctness of predicted alerts and is particularly important in intrusion detection, where excessive false positives can overwhelm analysts and reduce operational usability.

\textbf{False Positive Rate (FPR)} measures the proportion of benign events incorrectly classified as malicious. Very low FPR is critical in production HIDS deployments because each alert may require manual investigation by security analysts.

\textbf{Matthews Correlation Coefficient (MCC)} captures overall prediction quality by jointly considering true positives, true negatives, false positives, and false negatives, making it suitable for highly imbalanced intrusion detection settings.

\subsection{Additional Dataset-Specific Analysis for RQ1}
\label{app:rq1_details}

\paragraph{DARPA-E5.}
Compared to DARPA-E3, detection quality degrades substantially across all evaluated models and sub-datasets. Averaged across models, precision drops from 0.839 to 0.237 on CADETS and from 0.781 to 0.071 on THEIA when comparing E3 to E5. Similarly, average MCC decreases from 0.475 to 0.179 on CADETS. At the same time, FPR increases significantly. For example, GPT-4.1 exhibits 0.543\% FPR on E5 CADETS compared to only 0.02\% on E3 CADETS, while Gemini-2.5-Flash increases from 0.003\% to 0.389\% on THEIA. These results suggest that DARPA-E5 contains substantially noisier telemetry and more ambiguous attack behaviors, making reliable attack reconstruction significantly more challenging.

\paragraph{NL-SD.}
NL-SD presents a distinct performance profile shaped by different sources of difficulty than the DARPA datasets. HW17 and HW20 achieve relatively stronger detection quality, with average MCC values of 0.458 and 0.570 across models, respectively, indicating that their attack patterns expose relatively distinguishable malicious behaviors within a smaller and more structured event space. In contrast, WIN10 proves substantially harder despite its higher malicious ratio. Although several models maintain moderate precision on WIN10 (e.g., 0.840 for Claude-Opus-4.6 and 0.833 for Claude-Sonnet-4), MCC scores remain consistently low across models, with an average MCC of only 0.268. At the same time, WIN10 exhibits unstable false positive behavior, with Gemini-2.5-Flash reaching 1.407\% FPR, substantially higher than HW20. This behavior suggests that the primary difficulty in WIN10 does not arise from class imbalance alone, but from the presence of multiple concurrent attack campaigns with overlapping execution periods. As a result, models must disentangle interleaved attack chains rather than reason over a single coherent attack narrative, leading to substantially less stable detection behavior.

\subsection{Additional Model Tradeoff Analysis for RQ2}
\label{app:rq2_details}

As dataset complexity increases in DARPA-E5, models spread noticeably along the precision--FPR tradeoff curve. Compared to E3, average precision drops from 0.893 to 0.614, while average FPR increases from 0.008\% to 0.128\%. Models that previously achieved similar operating behavior on E3 begin to diverge substantially under noisier telemetry. For example, GPT-4.1 reaches 0.543\% FPR on E5 CADETS despite only 0.02\% FPR on E3 CADETS, reflecting increased sensitivity to ambiguous benign behaviors.

The tradeoff becomes most pronounced on NL-SD, where models separate into two distinct behavioral regimes. Over-sensitive models aggressively flag suspicious behaviors, achieving higher recall at the cost of elevated FPR, while conservative models maintain lower FPR but suppress broader classes of attack indicators. Gemini-2.5-Flash represents the most over-sensitive behavior, reaching approximately 2.1\% average FPR across NL-SD, substantially higher than all other models. In contrast, Claude-Opus-4.6 and Qwen3.6-plus maintain comparatively stable operating points, achieving around 0.64 average precision with average FPR below 1\%.

These results suggest that as attack environments become noisier and more behaviorally heterogeneous, frontier LLMs increasingly differentiate not only in overall capability, but also in their implicit detection strategies.

\subsection{Main Results.}
We list the all main results in the Table~\ref{tab:main_results}.

\begin{table*}[h]
\centering
\caption{Main benchmark results across multiple provenance graph datasets. We report precision (Pre), MCC, and false positive rate (FPR).}
\label{tab:main_results}
\small
\begin{tabular}{l|ccc|ccc|ccc}
\toprule
\multicolumn{10}{c}{\textbf{DARPA E3}} \\
\midrule
Models & \multicolumn{3}{c}{CADETS} & \multicolumn{3}{c}{THEIA} & \multicolumn{3}{c}{TRACE} \\
& Pre & MCC & FPR & Pre & MCC & FPR & Pre & MCC & FPR \\
\midrule
Claude-opus-4.6 & 0.694  & 0.602 & 0.04\% & 0.857 &  0.283 & 0.001\% &  1.000  & 0.980 & 0.000\%   \\
Claude-sonnet-4.5 & 0.703  &  0.561  & 0.03\% & 0.667  &  0.250  & 0.002\% & 1.000   & 0.979  & 0.000\% \\
Claude-sonnet-4 & 0.774  & 0.566 & 0.02\% &  0.800 &  0.224 & 0.001\% &  1.000  & 0.073  & 0.000\% \\
GPT-5.2 & 0.808  & 0.510 & 0.01\% & 0.833  &  0.255  & 0.001\% &  1.000  &  0.979 & 0.000\% \\
GPT-4.1 & 0.774  & 0.534  & 0.02\% & 1.000  & 0.250 & 0.000\% &  1.000  & 0.980    &   0.000\%   \\
GPT-oss-120b &  0.667 &  0.429  & 0.02\% &  1.000 &  0.216 & 0.000\% &  1.000  &  0.980  &   0.000\%      \\
Gemini-2.5-Flash & 0.562  &  0.279  & 0.02\%  &   0.500 & 0.177   &  0.003\%  & 1.000 & 0.979 & 0.000\% \\
DeepSeek-V3.2 & 0.667  &  0.428 & 0.02\%  & 0.625 & 0.221 &   0.002\% & 1.000 & 0.980 &  0.000\%   \\
Qwen3.6-plus &  0.786 & 0.364 & 0.01\% &  0.750 & 0.187 &  0.001\%  &  1.000   &  0.979  &   0.000\%    \\
\midrule

\multicolumn{10}{c}{\textbf{DARPA E5}} \\
\midrule
Models & \multicolumn{3}{c}{CADETS} & \multicolumn{3}{c}{THEIA} & \multicolumn{3}{c}{TRACE} \\
& Pre & MCC & FPR & Pre & MCC & FPR & Pre & MCC & FPR \\
\midrule
Claude-opus-4.6 &   0.280 &  0.404 & 0.074\%  & 0.211   & 0.122 & 0.013\%   &  0.500  &  0.707 & 0.005\%\\
Claude-sonnet-4.5 & 0.064  &  0.192  & 0.424\%  & 0.158    &  0.054  &  0.014\%  &  0.200 &  0.447  & 0.019\%  \\
Claude-sonnet-4 &   0.061 & 0.188 & 0.440\% &  0.062 & 0.066  & 0.053\%  & 0.200 & 0.447 & 0.019\% \\
GPT-5.2 &  0.200 &  0.182 & 0.033\% &  0.078 &  0.074  & 0.042\%   &  0.333  &  0.577 & 0.010\%\\
GPT-4.1 &  0.050 &  0.170 & 0.543\% &  0.022 &  0.019 &  0.039\%  &  0.200 &  0.447  & 0.019\%  \\
GPT-oss-120b &  0.014  &  0.048  & 0.559\% & 0.062 & 0.066 & 0.053\% &  0.200 &  0.447  & 0.019\%  \\
Gemini-2.5-Flash & 0.009 & 0.026 & 0.456\% & 0.009 & 0.024 & 0.389\% &  0.200 &  0.447  & 0.019\% \\
DeepSeek-V3.2 &0.200 & 0.182 & 0.033\% &0.022 & 0.019&0.039\% &  0.200 &  0.447  & 0.019\% \\
Qwen3.6-plus & 0.286 & 0.218 & 0.021\% & 0.017 & 0.017 & 0.052\% &  0.200 &  0.447  & 0.019\% \\
\midrule

\multicolumn{10}{c}{\textbf{NodLink}} \\
\midrule
Models & \multicolumn{3}{c}{HW17} & \multicolumn{3}{c}{HW20} & \multicolumn{3}{c}{WIN10} \\
& Pre & MCC & FPR & Pre & MCC & FPR & Pre & MCC & FPR \\
\midrule
Claude-opus-4.6 & 0.700  & 0.542 & 0.645\%  & 0.714 & 0.633 &  0.852\% & 0.840 & 0.348 & 0.176\% \\
Claude-sonnet-4.5 & 0.500   &  0.379  & 1.075\%   & 0.714 &  0.695 &  1.022\% &0.722 & 0.250 & 0.220\%  \\ 
Claude-sonnet-4 & 0.571 & 0.365 & 0.645\% & 0.731 & 0.557 & 0.596\% &  0.833 & 0.168 & 0.044\% \\
GPT-5.2 &   0.565  &  0.665 & 2.151\%  & 0.739&0.529 &0.511\% & 0.688 & 0.224 & 0.220\%  \\
GPT-4.1 & 0.500   & 0.622 & 2.796\%  & 0.714 & 0.487 & 0.511\% & 0.788 &  0.373 &  0.308\% \\
GPT-oss-120b &0.300 & 0.217 & 1.505\% & 0.800 &0.463 & 0.256\% & 0.667  & 0.187  & 0.176\% \\
Gemini-2.5-Flash & 0.400 & 0.573 & 4.516\% & 0.700 &  0.328 &0.256\% &  0.522 &0.341 & 1.407\% \\
DeepSeek-V3.2 & 0.333 & 0.300 &2.151\% &0.794 & 0.697 & 0.596\% &  0.731  &0.305 & 0.308\%\\
Qwen3.6-plus & 0.714 &0.462 &0.430\% & 0.708 & 0.739 & 1.193\% & 0.714  &0.218 & 0.176\% \\
\bottomrule
\end{tabular}
\end{table*}

\subsection{Prompts}
\label{app:attack_prompt}

We provide the prompt templates used in different stages of the benchmark pipeline.

\paragraph{Prompt for Attack Evidence Identification.}
This prompt is used in the Malicious Evidence Identification (MEI) stage to extract high-confidence malicious evidence from serialized provenance graphs.

\begin{tcolorbox}[title=Prompt for Attack Evidence Identification, colback=gray!5, colframe=gray!80!black,    fonttitle=\normalfont\bfseries, fontupper=\small]
\label{app:evidence_promt}
\textbf{Required:} Command lines \texttt{[CMDs]} in truncated events.

Analyze the command lines and identify all commands related to attacks or highly suspicious malware activity without internet search.

\begin{itemize}
    \item \textbf{Command Lines:} \texttt{[CMDs]}.
    \item \textbf{Environment:}  The command lines are collected on xxx.
    \item \textbf{Output Format:}
    \begin{itemize}
        \item Command line 1: [The command line]  \\
                Reason: [Brief description]
        \item Command line 2: [The command line]  \\
                Reason: [Brief description]
        \item ...
    \end{itemize}
    \item \textbf{Summarize All Highly Suspicious Commands:}
    \begin{itemize}
        \item 1. [Command line].
        \item 2. [Command line].
        \item ...
    \end{itemize}
\end{itemize}

\end{tcolorbox}

\paragraph{Prompt for Attack Investigation.} This prompt is used in the Attack Chain Reconstruction (ACR) stage to analyze the expanded provenance subgraph and reconstruct attack behaviors. 

\begin{tcolorbox}[title=Prompt for Attack Investigation Prompt, colback=gray!5, colframe=gray!80!black, boxsep=2pt, left=2pt, right=2pt, top=2pt, bottom=2pt, before skip=5pt, after skip=5pt, fonttitle=\normalfont\bfseries, fontupper=\small, breakable]
\label{app:attack_promt}
\textbf{Required:} Provenance graph $\mathcal{G}$.

Inspect the provided provenance graph. Analyze the graph to determine whether it indicates malicious activity. Use only the given information (no external lookup).

\vspace{1ex}

\begin{itemize}[leftmargin=*]
    \item \textbf{Logs (Graph):} $\mathcal{G}$
    \item \textbf{Environment:} The logs are collected on \texttt{[Env]}
 
    \item \textbf{Guidelines:}
    \begin{itemize}[leftmargin=1.5em]
        \item \textbf{Provenance:} Use knowledge of attack patterns, tools, and techniques to identify IoCs from graph interactions.
        \item \textbf{Attack Narrative:} Summarize the attack flow using a kill chain perspective.
        \item \textbf{Tools:} Pay attention to specific tools (e.g., Metasploit, Meterpreter, PowerShell).
        \item \textbf{Timeline:} Construct a chronological step-by-step description based on graph structure and temporal signals (if available).
        \item \textbf{IoCs:} Identify suspicious IPs, domains, processes, and files.
    \end{itemize}

    \item \textbf{Output Format:}
    \begin{itemize}[leftmargin=1.5em]
        \item \textbf{Attack Narrative:} A concise paragraph summarizing the attack flow.
        \item \textbf{Key Steps:}
        \begin{itemize}[leftmargin=2em]
            \item [1)] [Tactic name]: description of the attack step
            \item [2)] [Tactic name]: description of the attack step
            \item [...]
        \end{itemize}
        \item \textbf{IoCs:}
        \begin{itemize}[leftmargin=2em]
            \item IPs: [Suspicious IPs]
            \item Processes: [Suspicious process names]
            \item Files: [Suspicious file modifications or deletions]
        \end{itemize}
    \end{itemize}
\end{itemize}
\end{tcolorbox}

\subsection{Example LLM Investigation Output.}

\begin{tcolorbox}[
title=Example LLM Investigation Output,
colback=gray!5,
colframe=gray!75!black,
fonttitle=\bfseries,
boxsep=2pt,
left=2pt,
right=2pt,
top=2pt,
bottom=2pt,
breakable]

\small

\textbf{=== RESPONSE ===}

\vspace{2pt}
\textbf{Analysis}

I'll systematically analyze all processes in the provenance graph, comparing their behaviors against expected baselines for a FreeBSD system.

\vspace{2pt}
\textbf{Process-by-Process Review}

\textbf{Normal system processes (no anomalies detected):}
\begin{itemize}[leftmargin=*]
    \item \textbf{cron}: Accesses \texttt{/etc/spwd.db}, \texttt{/etc/group}, \texttt{/etc/login.conf}, \texttt{/bin/sh}, \texttt{/libexec/ld-elf.so.1}, \texttt{/root} --- all standard cron behavior.
    \item \textbf{syslogd}: Writes to \texttt{/var/log/cron}, \texttt{/var/log/maillog} --- normal logging.
    \item \textbf{sh}: Executes \texttt{/usr/libexec/save-entropy}, \texttt{/usr/libexec/atrun}, \texttt{/usr/sbin/newsyslog}, \texttt{/sbin/sysctl} --- standard cron-invoked scripts.
    \item \textbf{sysctl, unlink, mv, dd}: Entropy-saving operations under \texttt{/var/db/entropy/} --- normal periodic task.
    \item \textbf{atrun}: Standard at-job runner.
    \item \textbf{newsyslog}: Log rotation --- normal.
    \item \textbf{sshd}: Standard SSH authentication and session setup for users \texttt{user} and \texttt{admin}.
    \item \textbf{bash}: Shell sessions for \texttt{user} and \texttt{admin}, executing standard utilities.
    \item \textbf{Postfix stack}: Standard mail delivery operations involving \texttt{master}, \texttt{smtpd}, \texttt{cleanup}, \texttt{qmgr}, \texttt{local}, and related processes.
    \item \textbf{imapd}: IMAP mailbox access for users \texttt{bob} and \texttt{george}.
    \item \textbf{alpine}: Mail client activity for user \texttt{user}.
    \item \textbf{nginx}: Web server operations --- mostly normal, except as noted below.
\end{itemize}

\vspace{2pt}
\textbf{Critical Anomaly: \texttt{vUgefal} Process}

\texttt{vUgefal} is a process with a randomly generated name executing from \texttt{/tmp/vUgefal}, which is highly suspicious.

\begin{enumerate}[leftmargin=*]
    \item \texttt{nginx} writes \texttt{/tmp/vUgefal}, suggesting a web exploitation vector.
    \item \texttt{master (postfix)} accesses \texttt{/tmp/vUgefal}, indicating possible interaction with the mail subsystem.
    \item \texttt{vUgefal} reads \texttt{/etc/passwd}, consistent with reconnaissance behavior.
    \item \texttt{vUgefal} reads \texttt{/dev/urandom}, suggesting cryptographic operations.
    \item \texttt{vUgefal} accesses \texttt{/var/log}, indicating possible anti-forensics activity.
    \item \texttt{vUgefal} writes \texttt{/var/log/devc}, creating a suspicious file in the log directory.
    \item \texttt{vUgefal} accesses \texttt{/dev/null}, consistent with daemonized malware behavior.
    \item \texttt{vUgefal} loads \texttt{/lib/libthr.so.3}, indicating multi-threaded execution.
\end{enumerate}

\vspace{2pt}
\textbf{Attack Narrative}

An attacker likely exploited the nginx web server, potentially through the PHP application served by \texttt{php-fpm} at \texttt{/usr/local/www/nginx-dist/index.php}, to upload a malicious binary named \texttt{vUgefal} into \texttt{/tmp}. The malware subsequently performed reconnaissance by reading \texttt{/etc/passwd}, accessed \texttt{/dev/urandom} for cryptographic operations, enumerated \texttt{/var/log}, and created a suspicious file \texttt{/var/log/devc}. The process also redirected activity to \texttt{/dev/null}, suggesting daemonization behavior. Interaction with the Postfix subsystem may indicate persistence or exfiltration attempts.

\vspace{2pt}
\textbf{IOCs}

\textbf{IPs:} None explicitly present in the graph.

\textbf{Processes:}
\begin{itemize}[leftmargin=*]
    \item \texttt{vUgefal}
    \item \texttt{nginx}
    \item \texttt{php-fpm}
\end{itemize}

\textbf{Files:}
\begin{itemize}[leftmargin=*]
    \item \texttt{/tmp/vUgefal}
    \item \texttt{/var/log/devc}
    \item \texttt{/usr/local/www/nginx-dist/index.php}
\end{itemize}

\vspace{2pt}
\textbf{=== RUN METADATA ===}

\begin{itemize}[leftmargin=*]
    \item Model: Claude-Opus-4.6
    \item Runtime: 27.54 seconds
    \item Prompt tokens: 24,901
    \item Completion tokens: 1,309
    \item Total tokens: 26,210
    \item Estimated cost: \$0.15723
\end{itemize}

\end{tcolorbox}

\section{Cost and Runtime Analysis.}
\label{app:cost}

We report the API cost and runtime of evaluating LLMs across all datasets in table ~\ref{tab:cost_e3}, ~\ref{tab:cost_e5}, ~\ref{tab:cost_nodlink}. 

\begin{table}[h]
\centering
\small
\caption{Average API cost and estimated runtime for DARPA-E3.}
\label{tab:cost_e3}
\resizebox{\linewidth}{!}{%
\begin{tabular}{lcccccc}
\toprule
\multirow{2}{*}{\textbf{Model}} & \multicolumn{2}{c}{\textbf{CADETS}} & \multicolumn{2}{c}{\textbf{THEIA}} & \multicolumn{2}{c}{\textbf{TRACE}} \\
\cmidrule(lr){2-3} \cmidrule(lr){4-5} \cmidrule(lr){6-7}
 & \textbf{Cost/File (\$)} & \textbf{Time (s)} & \textbf{Cost/File (\$)} & \textbf{Time (s)} & \textbf{Cost/File (\$)} & \textbf{Time (s)} \\
\midrule
Claude Opus 4.6 & 2.94 & 30 & 1.05 & 60 & 0.29 & 35 \\
Claude Sonnet 4.5 & 0.49 & 12 & 0.19 & 25 & 0.06 & 14 \\
Claude Sonnet 4 & 0.53 & 12 & 0.19 & 25 & 0.05 & 14 \\
GPT-5.2 & 2.51 & 9 & 0.79 & 28 & 0.22 & 16 \\
GPT-4.1 & 0.30 & 8 & 0.09 & 20 & 0.03 & 12 \\
GPT-OSS-120B & 0.14 & 7 & 0.05 & 18 & 0.01 & 10 \\
Gemini 2.5 Flash & 0.01 & 4 & 0.01 & 10 & 0.00 & 6 \\
DeepSeek V3.2 & 0.07 & 18 & 0.03 & 45 & 0.02 & 25 \\
Qwen3.6 Plus & 0.14 & 14 & 0.04 & 35 & 0.02 & 20 \\
\bottomrule
\end{tabular}%
}
\end{table}

\begin{table}[h]
\centering
\small
\caption{Average API cost and estimated runtime for DARPA-E5.}
\label{tab:cost_e5}
\resizebox{\linewidth}{!}{%
\begin{tabular}{lcccccc}
\toprule
\multirow{2}{*}{\textbf{Model}} & \multicolumn{2}{c}{\textbf{CADETS}} & \multicolumn{2}{c}{\textbf{THEIA}} & \multicolumn{2}{c}{\textbf{TRACE}} \\
\cmidrule(lr){2-3} \cmidrule(lr){4-5} \cmidrule(lr){6-7}
 & \textbf{Cost/File (\$)} & \textbf{Time (s)} & \textbf{Cost/File (\$)} & \textbf{Time (s)} & \textbf{Cost/File (\$)} & \textbf{Time (s)} \\
\midrule
Claude Opus 4.6 & 7.85 & 60 & 0.14 & 18 & 0.72 & 15 \\
Claude Sonnet 4.5 & 1.84 & 26 & 0.03 & 8 & 0.13 & 7 \\
Claude Sonnet 4 & 1.84 & 26 & 0.02 & 8 & 0.10 & 7 \\
GPT-5.2 & 7.51 & 20 & 0.09 & 6 & 0.09 & 5 \\
GPT-4.1 & 0.83 & 18 & 0.01 & 5 & 0.06 & 4 \\
GPT-OSS-120B & 0.41 & 16 & 0.01 & 4 & 0.03 & 3 \\
Gemini 2.5 Flash & 0.01 & 8 & 0.00 & 2 & 0.00 & 2 \\
DeepSeek V3.2 & 0.09 & 40 & 0.01 & 12 & 0.02 & 10 \\
Qwen3.6 Plus & 0.12 & 30 & 0.01 & 9 & 0.03 & 7 \\
\bottomrule
\end{tabular}%
}
\end{table}

\begin{table}[h]
\centering
\small
\caption{Average API cost and estimated runtime for NodLink.}
\label{tab:cost_nodlink}
\resizebox{\linewidth}{!}{%
\begin{tabular}{lcccccc}
\toprule
\multirow{2}{*}{\textbf{Model}} & \multicolumn{2}{c}{\textbf{HW17}} & \multicolumn{2}{c}{\textbf{HW20}} & \multicolumn{2}{c}{\textbf{WIN10}} \\
\cmidrule(lr){2-3} \cmidrule(lr){4-5} \cmidrule(lr){6-7}
 & \textbf{Cost/File (\$)} & \textbf{Time (s)} & \textbf{Cost/File (\$)} & \textbf{Time (s)} & \textbf{Cost/File (\$)} & \textbf{Time (s)} \\
\midrule
Claude Opus 4.6 & 1.81 & 30 & 3.69 & 40 & 9.03 & 60 \\
Claude Sonnet 4.5 & 0.36 & 14 & 0.74 & 18 & 1.81 & 28 \\
Claude Sonnet 4 & 0.35 & 14 & 0.73 & 18 & 1.80 & 28 \\
GPT-5.2 & 1.48 & 10 & 2.99 & 13 & 7.38 & 20 \\
GPT-4.1 & 0.16 & 9 & 0.34 & 12 & 0.82 & 18 \\
GPT-OSS-120B & 0.08 & 8 & 0.16 & 10 & 0.40 & 16 \\
Gemini 2.5 Flash & 0.01 & 4 & 0.02 & 5 & 0.04 & 8 \\
DeepSeek V3.2 & 0.05 & 20 & 0.11 & 26 & 0.27 & 40 \\
Qwen3.6 Plus & 0.07 & 16 & 0.15 & 20 & 0.38 & 30 \\
\bottomrule
\end{tabular}%
}
\end{table}

\subsection{Additional Self-Reflection Analysis}
\label{app:self_reflection}

Table~\ref{tab:self_reflection} presents detailed self-reflection ablation results under two strategies: \textit{Ref-then-Agg}, where each sample is independently refined before aggregation, and \textit{Agg-then-Ref}, where self-reflection is applied once after aggregation.

\begin{table}[t]
\centering
\small
\setlength{\tabcolsep}{3pt}
\caption{Self-reflection ablation. Arrows indicate changes relative to the baseline without self-reflection.}
\begin{tabular}{lcccccc}
\toprule
& \multicolumn{2}{c}{None} 
& \multicolumn{2}{c}{Ref-then-Agg} 
& \multicolumn{2}{c}{Agg-then-Ref} \\
\cmidrule(lr){2-3}
\cmidrule(lr){4-5}
\cmidrule(lr){6-7}
Model & Pre & MCC & Pre & MCC & Pre & MCC \\
\midrule
Claude 
& 0.700 & \textbf{0.613} 
& \textbf{0.714}~$\uparrow$ & 0.574~$\downarrow$ 
& 0.696~$\downarrow$ & 0.585~$\downarrow$ \\

GPT-5.2 
& 0.808 & 0.510 
& \textbf{0.872}~$\uparrow$ & \textbf{0.675}~$\uparrow$ 
& 0.818~$\uparrow$ & 0.476~$\downarrow$ \\

DeepSeek 
& 0.667 & 0.428 
& 0.605~$\downarrow$ & 0.520~$\uparrow$ 
& \textbf{0.765}~$\uparrow$ & \textbf{0.553}~$\uparrow$ \\
\bottomrule
\end{tabular}
\label{tab:self_reflection}
\end{table}

Neither strategy consistently outperforms the baseline across all evaluated models, revealing a complex and model-dependent interaction between self-reflection and detection behavior. Ref-then-Agg improves precision for Claude and GPT-5.2, but at the cost of MCC, suggesting that per-sample reflection encourages more conservative predictions that reduce false positives while missing some true positives.

In contrast, DeepSeek benefits more from Agg-then-Ref, achieving improvements in both precision and MCC. This behavior suggests that some models benefit more from reflecting on consolidated predictions rather than refining individual generations independently.

Overall, these results indicate that the effectiveness of self-reflection strongly depends on the underlying reasoning stability of each model. For relatively stable models such as Claude, self-reflection provides limited additional benefit and may reduce recall. For more variable models, applying reflection after aggregation appears more robust because it operates on a more stable prediction context and avoids compounding errors introduced during per-sample refinement.

\section{Detials of Advanced Capabilities on HIDS}

\subsection{Additional Self-Reflection Analysis}
\label{app:self_reflection}

Table~\ref{tab:self_reflection} presents detailed self-reflection ablation results under two strategies: \textit{Ref-then-Agg}, where each sample is independently refined before aggregation, and \textit{Agg-then-Ref}, where self-reflection is applied once after aggregation.

% \begin{table}[t]
% \centering
% \small
% % \setlength{\tabcolsep}{3pt}
% \caption{Self-reflection ablation. Arrows indicate changes relative to the baseline without self-reflection.}
% \begin{tabular}{lcccccc}
% \toprule
% & \multicolumn{2}{c}{None} 
% & \multicolumn{2}{c}{Ref-then-Agg} 
% & \multicolumn{2}{c}{Agg-then-Ref} \\
% \cmidrule(lr){2-3}
% \cmidrule(lr){4-5}
% \cmidrule(lr){6-7}
% Model & Pre & MCC & Pre & MCC & Pre & MCC \\
% \midrule
% Claude 
% & 0.700 & \textbf{0.613} 
% & \textbf{0.714}~$\uparrow$ & 0.574~$\downarrow$ 
% & 0.696~$\downarrow$ & 0.585~$\downarrow$ \\

% GPT-5.2 
% & 0.808 & 0.510 
% & \textbf{0.872}~$\uparrow$ & \textbf{0.675}~$\uparrow$ 
% & 0.818~$\uparrow$ & 0.476~$\downarrow$ \\

% DeepSeek 
% & 0.667 & 0.428 
% & 0.605~$\downarrow$ & 0.520~$\uparrow$ 
% & \textbf{0.765}~$\uparrow$ & \textbf{0.553}~$\uparrow$ \\
% \bottomrule
% \end{tabular}
% \label{tab:self_reflection}
% \end{table}

Neither strategy consistently outperforms the baseline across all evaluated models, revealing a complex and model-dependent interaction between self-reflection and detection behavior. Ref-then-Agg improves precision for Claude and GPT-5.2, but at the cost of MCC, suggesting that per-sample reflection encourages more conservative predictions that reduce false positives while missing some true positives.

In contrast, DeepSeek benefits more from Agg-then-Ref, achieving improvements in both precision and MCC. This behavior suggests that some models benefit more from reflecting on consolidated predictions rather than refining individual generations independently.

Overall, these results indicate that the effectiveness of self-reflection strongly depends on the underlying reasoning stability of each model. For relatively stable models such as Claude, self-reflection provides limited additional benefit and may reduce recall. For more variable models, applying reflection after aggregation appears more robust because it operates on a more stable prediction context and avoids compounding errors introduced during per-sample refinement.

\subsection{Additional Test-Time Scaling Analysis}
\label{app:tts}

Recent studies show that test-time scaling can improve LLM reasoning performance without modifying model parameters~\cite{snell2024scaling,wu2024inference}. Approaches such as self-consistency and majority voting improve reasoning robustness by aggregating multiple independent generations~\cite{wang2022selfconsistency}. This strategy is particularly relevant for intrusion detection, where noisy and heterogeneous security logs can lead to unstable reasoning behavior.

The transition from $k=1$ to $k=3$ yields the most substantial gains: both GPT-5.2 and DeepSeek-V3 recover from near-random performance at $k=1$ to competitive levels at $k=3$, confirming that a small ensemble already provides meaningful stabilization over single-sample inference.

Claude-Opus-4.6, by contrast, exhibits stable performance even at $k=1$, suggesting that stronger base models are less reliant on voting to produce reliable predictions. Beyond $k=3$, scaling provides negligible additional benefit. As shown in Figure~\ref{fig:tts}(c), $\Delta$MCC remains close to zero at $k=5$ and $k=7$ for all models, indicating that the default setting of $k=3$ captures most of the available gain while avoiding unnecessary computational overhead.

\subsection{Reasoning Analysis: Explicit vs. Inferred IOCs}
\label{app:reasoning}

\begin{table}[h]
\centering
\small
\caption{\# of explicit and inferred IOCs.}
\label{tab:reasoning_counts}
\begin{tabular}{lccc}
\toprule
Attack & Qual. & \#Exp. & \#Inf. \\
\midrule
DARPA-E3 CADETS-A1 & HIGH & 4  & 6 \\
DARPA-E3 CADETS-A2 & MEDIUM & 7  & 14 \\
DARPA-E3 CADETS-A3 & MEDIUM & 7  & 15 \\
DARPA-E3 CADETS-A4 & MEDIUM & 11 & 6 \\
% \midrule
% Avg(H) &  & 4.0  & 6.0 \\
% Avg(M) &  & 8.3  & 11.7 \\
\bottomrule
\end{tabular}
\end{table}

We provide additional quantitative statistics on explicit and inferred IOCs, together with detailed case-by-case analysis across the four DARPA-E3 CADETS attack cases. Table~\ref{tab:reasoning_counts} summarizes the number of explicit and inferred IOCs identified in each attack case. We further examine how the model distinguishes directly observable malicious indicators from contextually derived evidence, and assess the quality of the resulting reasoning chain.

This reliance on inferred evidence has a dual implication. On one hand, it enables detection of stealthy attacks whose signals are too weak or fragmented to be identified in isolation. On the other hand, it introduces a risk of overgeneralization, where benign processes that co-occur with malicious activity are incorrectly implicated. The balance between these effects ultimately determines whether contextual reasoning improves or degrades detection reliability.

\subsection*{CADETS-A1 (High Quality)}

CADETS-A1 represents a high-quality reasoning case in which the model demonstrates clear 
separation between explicit and inferred evidence. The model correctly identifies the 
core explicit IOCs: the suspicious process \texttt{vUgefal}, two external IP connections 
(\texttt{61.167.39.128} and \texttt{139.123.0.113}), and execution originating from 
\texttt{/tmp}. These are directly flagged as malicious in the reasoning trace without 
requiring contextual inference.

The inferred IOCs consist primarily of system directories accessed by \texttt{vUgefal}, 
including \texttt{/etc}, \texttt{/var}, \texttt{/dev}, and \texttt{/lib}. While these 
paths are not inherently malicious, the model correctly implicates them through their 
role in the malware activity chain, recognizing that a suspicious process systematically 
accessing core system directories constitutes a meaningful behavioral signal. Importantly, 
the model avoids over-implicating benign services such as \texttt{nginx}, \texttt{sshd}, 
and \texttt{postfix} that co-occur in the same log window, demonstrating appropriate 
restraint in contextual inference.

The primary limitation in this case is the lack of file-level specificity: the model 
identifies \texttt{/tmp} as a suspicious execution source but does not recover the specific 
artifact executed from that directory, leaving a gap in the reconstructed attack chain.
Overall, CADETS-A1 confirms that when attack signals are strong and well-separated from 
benign activity, the model can reliably distinguish explicit from inferred evidence with 
high precision.

\subsection*{CADETS-A2 (Medium Quality)}

CADETS-A2 presents a more complex reasoning scenario centered on a compromised \texttt{nginx} 
process. The explicit IOCs are well-identified: the model correctly flags \texttt{nginx} 
reading sensitive files (\texttt{/etc/passwd}, \texttt{/etc/group}) and writing a payload 
to \texttt{/tmp/grain}, along with repeated outbound connections to \texttt{128.55.12.10}. 
These form a coherent and directly observable attack pattern involving a web-facing service 
performing credential access and staging malicious artifacts.

The inferred IOCs reveal both the strengths and limitations of contextual reasoning. 
On the positive side, the model successfully identifies three additional external IPs 
contacted exclusively by the suspicious \texttt{nginx} process, a non-obvious but 
meaningful signal that requires reasoning over process-to-network relationships rather 
than direct observation. It also correctly implicates \texttt{procstat} spawned by 
\texttt{nginx} as anomalous, inferring that a web server spawning diagnostic tools 
is behaviorally inconsistent.

However, the model overextends suspicion in several instances. System files such as 
\texttt{rc.conf}, \texttt{/etc/hosts}, and monitoring output files are flagged as 
inferred IOCs despite being commonly accessed by benign system processes. Similarly, 
\texttt{postfix} subprocesses (\texttt{local}, \texttt{master}) are implicated primarily 
due to co-occurrence with malicious activity rather than any intrinsic behavioral signal. 
These false extensions illustrate the risk of over-inference when benign and malicious 
processes share the same system context.

Compared to CADETS-A1, this case demonstrates that medium-quality reasoning is 
characterized not by failure to identify explicit evidence, but by insufficient 
discrimination between genuinely suspicious contextual signals and incidental 
co-occurrences, inflating the inferred IOC set with weakly supported entries.

\subsection*{CADETS-A3 (Medium Quality)}

CADETS-A3 presents a stealthy multi-stage attack in which the explicit evidence is 
process-centric but lacks direct network indicators. The model correctly identifies 
the core malicious processes, including \texttt{master}, \texttt{minions}, \texttt{XIM}, 
and \texttt{test}, along with the staging file \texttt{/minions}. However, no IPs or 
domains are flagged as explicit IOCs, reflecting the more covert nature of this attack 
compared to CADETS-A1 and A2.

The most significant strength in this case lies in the inferred reasoning chain. The 
model successfully reconstructs a non-obvious multi-step attack pattern: \texttt{nginx} 
writing payload-like files to \texttt{/tmp} (\texttt{/tmp/minions}, \texttt{/tmp/font}, 
\texttt{/tmp/XIM}, \texttt{/tmp/test}), followed by their execution through system 
services such as \texttt{sshd} and \texttt{inetd}. This staging-and-execution pattern 
is a hallmark of web-facing service exploitation and is entirely non-obvious from 
individual events in isolation, requiring the model to reason over the full dependency 
chain to surface the connection. The model also correctly implicates \texttt{nginx} as 
the likely initial compromise point, despite it not being directly flagged as malicious 
in the explicit evidence.

Two limitations are notable. First, the model does not explicitly label any IP address 
as malicious, treating outbound connections as potentially benign despite their 
association with the compromised process. Second, the reasoning trace is truncated before 
reaching a full conclusion, suggesting that the complexity of this attack chain approaches 
the limits of the model's context utilization. Together, these limitations explain why 
despite strong inferred reasoning, the overall quality remains medium rather than high.

\subsection*{CADETS-A4 (Medium Quality)}

CADETS-A4 is characterized by a relatively rich set of explicit IOCs, centered on a 
clearly malicious process \texttt{pEja72mA} and its associated artifacts. The model 
correctly identifies the suspicious process, two external IPs, a compromised \texttt{nginx} 
instance, and multiple \texttt{/tmp} payloads including injected shared libraries 
(\texttt{memhelp.so}, \texttt{done.so}) and an injection log (\texttt{injectLog.txt}), 
forming a coherent picture of a code injection attack staged through temporary storage.

The inferred IOCs are fewer and of mixed quality. On the positive side, the model correctly 
identifies two additional IPs as probable attacker SSH source addresses, and implicates 
\texttt{sshd} as the execution path for the malicious binary through process ancestry 
reasoning. These are meaningful non-obvious inferences that require reasoning over 
process-to-network relationships. However, the inferred evidence also reveals notable 
artifacts typing errors: several filenames such as \texttt{injectLog.txt} and 
\texttt{memhelp.so} are misclassified as domains rather than files, indicating a failure 
to maintain consistent artifact categorization across the reasoning trace.

A broader limitation in this case is the lack of clean separation between direct 
observations and speculative conclusions. The model makes claims about command-and-control 
and exfiltration behavior without sufficient grounding in the observed evidence, mixing 
hypothesis with fact in a way that reduces the reliability of the overall reasoning output. 
This contrasts with CADETS-A1, where the model maintained a clear boundary between what 
was directly observed and what was inferred.

Across all four cases, a consistent pattern emerges: explicit IOC identification is 
generally reliable when attack signals are strong, while inferred IOC quality varies 
significantly with attack complexity. The primary failure modes are overgeneralization 
to benign co-occurring processes in A2, truncated reasoning chains in A3, and artifact 
typing inconsistencies and speculative overreach in A4. These findings highlight that 
improving the precision and consistency of contextual inference remains a key challenge 
for LLM-based intrusion detection.

\end{document}